**EIC (Expert Information Criterion) not AIC:**

**The cautious biologist's guide to model selection and causal inference**

Zachary M. Laubach[1,2], Eleanor J. Murray[3], Kim L. Hoke[4], Rebecca J. Safran[1], Wei Perng[5]

[1] Department of Ecology and Evolutionary Biology, University of Colorado, Boulder, CO, USA
[2] Department of Integrative Biology, Michigan State University, Lansing, MI, USA
[3] Department of Epidemiology, Boston University School of Public Health, Boston, MA, USA
[4] Department of Biology, Colorado State University, Fort Collins, CO, USA
[5] Department of Epidemiology, University of Colorado Denver Anschutz Medical Campus, Aurora, CO, USA

**Correspondence**:
Zachary Laubach (Zachary.Laubach@colorado.edu)

**ABSTRACT**

1. A goal of many research programs in biology is to extract meaningful insights from large, complex data sets. Researchers in Ecology, Evolution and Behavior (EEB) often grapple with long-term, observational data sets from which they construct models to address fundamental questions about biology. Similarly, epidemiologists analyze large, complex observational data sets to understand the distribution and determinants of human health and disease. A key difference in the analytical workflows for these two distinct areas of biology is delineation of data analysis tasks and explicit use of causal inference methods, widely adopted by epidemiologists.
2. Here, we review the most recent causal inference literature and describe an analytical workflow that has direct applications for EEB researchers.
3. The first half of this commentary defines four distinct analytical tasks (description, prediction, association, and causal inference), and the corresponding approaches to data analysis and model selection. The latter half is dedicated to walking the reader through the steps of casual inference, focusing on examples from EEB.
4. Given increasing interest in causal inference and common misperceptions regarding the task of causal inference, we aim to facilitate an exchange of ideas between disciplinary silos and provide a framework for analyses of all data, though particularly relevant for observational data.

## 1. INTRODUCTION

In 1976, George Box noted, “All models are wrong.” (Box, 1976). While this aphorism is true, models are necessary to summarize data and draw inference about the world. As such, the process of specifying, testing, and implementing a model cannot be decoupled from advancing knowledge, developing theory, and crafting policy. Thus, researchers find themselves between the Scylla of accepting a model’s shortcomings and the Charybdis that is the foundational role of models in evidence-based science.

In this commentary, we frame the process of model selection around asking the right questions. This feat involves identifying the appropriate analytical task and aligning the modeling approach with the task so that the model, though not perfectly parameterized, improves our understanding of biology. All scientists benefit from a systematic approach to data analysis, but the target audience for this commentary is scientists in the Ecology, Evolution and Behavior (EEB) discipline – a field that tests hypotheses and draws conclusions from observational data collected from wild organisms in pursuit of proximate and ultimate explanations of the natural world (Tinbergen, 1963). While often unstated, a key goal of EEB is to infer causality from observational data that are vulnerable to extraneous variability and systematic error. Analyses of such data require careful consideration of covariates in multiple variable analyses, whereas such issues are of less (but not no) concern for scientists interrogating data generated from controlled laboratory experiments. EEB scientists have extensive quantitative training, so giving advice on technicalities of statistical methods is not the goal of this paper. Rather, we share a systematic way of thinking about data analysis in order to obtain the appropriate answer to a clearly defined research question.

The first half of this paper links specific and clear research questions to the type of analytical task, followed by suggestions on example methods/models for each task. In the second half, we focus specifically on causal inference. Though not new in EEB (Wright, 1921; Shipley, 1999), causal inference has experienced a recent resurgence of interest and is highly relevant to the types of data collected and analyzed by behavioral ecologists and evolutionary biologists (Larsen, Meng, & Kendall, 2019; Laubach et al., 2020; Rosenbaum et al., 2020; Zeng, Rosenbaum, Archie, Alberts, & Li, 2020).

## 2. ASKING THE RIGHT QUESTION

A researcher needs to collect blood from 100 animals and store it in a lysis buffer for lab assays. Recalling that her colleague recently collected blood for this exact purpose, she asks her colleague, "How much buffer did you purchase for 100 animals?" The colleague replies, "Two 100mL bottles." Following this advice, the researcher buys two 100mL bottles of buffer and was left with an unused bottle after the job was complete. Vexed by the extra expense, the researcher inquires of her colleague, "Why did you not tell me that a single 100mL bottle of lysis buffer was needed?" To which the colleague replies, "You asked, how much buffer I *bought*, not how much I *used*."

This facetious example exemplifies the importance of asking the right question. Nature, represented by raw data, provides an unbiased truth (Pearl & Mackenzie, 2018). Extracting the truth we seek from the data we analyze depends on the question we ask. A real danger lies in a scenario where we unknowingly ask the wrong question, but we interpret the answer in the context of the question we meant to ask rather than the one posed.

## 3. DATA ANALYSIS TASKS

Identifying the appropriate analytical task for a research question is the first step to asking the right question. Here, we describe four discrete tasks: description, prediction, association, and causal inference (**Box 1**), identified through a review of the most recent causal inference literature (Hernán, Hsu, & Healy, 2019; Conroy & Murray, 2020). In addition to defining each task, we discuss when use of each task is appropriate, the types of question that each can answer, and how combinations of tasks may be coordinated and used together in single manuscript or to understand a specific aspect of biology.

**Box 1.** Data analysis tasks

| **Task** | **Key characteristics and concepts** | **Analytical tools** | **Causal knowledge needed?** | **Example research question** |
|---|---|---|---|---|
| Description | A quantitative summary of the data. The metrics of interest may range from simple descriptive statistics to complex visualization techniques. | Means ± SD, box plots, proportions, unsupervised cluster analyses, time trends | No | What is the central tendency and spread of T-cell count, a marker of immune function, in wild spotted hyenas in Kenya? |
| Prediction | Identification of a set of explanatory variables that optimize variation explained in a dependent variable, with no focus on the causal or temporal structure among the explanatory variables of interest. This task often involves use of automated procedures to maximize model fit and leverages the joint distribution of multiple variables. | Tree-based techniques; recurrent neural networks; unsupervised machine learning algorithms | No | What set of social and ecological factors explain maximum variation in T-cell count in wild spotted hyenas in Kenya? |
| Association | Assessment of the unadjusted relationship between two variables of interest. This relationship may be explored within strata of a few key other variables that may influence the association of interest and can inform downstream causal inference analysis. | Pearson or Spearman correlation coefficients, estimates from unadjusted regression models | Some | How does social connectedness correlate with T-cell count in wild spotted hyenas in Kenya? |
| Causal inference | The goal is to obtain an unbiased effect of an explanatory variable(s) of interest (X) on a dependent variable (Y). This type of analysis requires expert knowledge on the causal and temporal relationship between X and Y, as well as additional third variables (confounders, mediators, effect modifiers, colliders) that may influence this relationship in order to properly parameterize statistical models. | Use of directed acyclic graphs to reflect a specific research question and summarize the analytical technique, followed by an appropriate systematic approach. | Yes | Does social connectedness affect T-cell count in wild spotted hyenas in Kenya? |

### 3.1. Description: "What do X and/or Y look like?"

Descriptive analyses provide an unbiased overview of the data. The techniques may be simple calculations of central tendency, variation, or frequency. They can also be complex and involve unsupervised machine learning techniques to visualize the data. Such analyses are valuable for elucidating the probability function(s) that generate the data and for identifying naturally-occurring trends and group structure in the data (Holmes & Huber, 2019). In EEB, a familiar example of descriptive analyses involves characterizing the size of a population of animals over time (Green, Johnson-Ulrich, Couraud, & Holekamp, 2018).

### 3.2. Prediction: "Which combination of X-variables optimize variability explained in Y?"

The goal of prediction is to build a model that provides accurate quantitative guesses at the value of the Y-variable for a given set of X-variables. This task often relies on metrics of model fit such as $R^2$ values or Akaike information criterion (AIC) and Bayesian Information Criterion (BIC), and tests of statistical significance to arrive at the final model. Techniques for this task include tree-based methods, neural networks, and machine learning algorithms (Holmes & Huber, 2019; Hothorn, 2020). An example of prediction from EEB is identifying ecological factors, like prey abundance, that explain variation in predator abundance in test populations, and then assessing how well prey abundance forecasts predator abundance in external populations (Karanth, Nichols, Kumar, Link, & Hines, 2004).

For this task, the interrelations among the X-variables are not of interest, and the investigator should not interpret estimates for individual X-variables. Many researchers seeking to make causal inference erroneously take a predictive modeling approach due, in part, to how introductory statistics courses teach model selection (i.e., use of forward or backward selection algorithms to arrive at a final set of covariates). This is among the most common reasons for why the answer obtained from an analysis does not reflect the original inquiry, an issue that transpires from discrepant temporal and causal relationships among X-variables included as predictors – a topic we revisit in the causal inference section. Another often-overlooked aspect of prediction is a practical one: true prediction requires validation (Perng & Aslibekyan, 2020). The value of a predictive model is judged by its capability to accurately predict Y in an independent set of data, in addition to how well the model fits the original training dataset. Given the complexity of many outcomes in real-world settings, prediction is a tall task that warrants separate discussion beyond the scope of this commentary (see: (Mac Nally, 2000; Mac Nally, Duncan, Thomson, & Yen, 2018) for more information).

**3.3. Association: "If X occurs, can we expect Y to occur? How are X and Y related?"**

Associational analyses provide an understanding of the crude relationship between two variables – either by themselves, or within strata of a few key variables that are identified *a priori.* As with prediction, one should not interpret the parameter estimate with causality in mind since associational analyses do not seek to attribute an *effect* of one variable on another. For instance, an investigator may be interested in the relationship between carrying a lighter and risk of lung cancer. The associational analysis reveals that carrying a lighter is associated with risk of lung cancer. While this relationship is obviously not causal – i.e., there is no biological knowledge suggesting that having a lighter in your pocket can trigger oncogenic pathways – such findings are useful for informing future studies of causal inference and can inform prediction models by identifying a set of variables that account for maximal variation in an outcome. The investigator may use this information in conjunction with *a priori* knowledge that lighters are necessary to smoke carcinogenic tobacco products, and subsequently formulate a causal research question on the effect of cigarette smoking on lung cancer. An example from EEB is an older study that assessed the crude associations between key behavioral and demographic characteristics of farm workers in relation to *Toxoplasma gondii* infection, which serves as a foundation for future studies to home in on causal risk factors of infection (Weigel, Dubey, Dyer, & Siegel, 1999).

### 3.4. Causal inference: "Does X cause Y?"

The goal of causal inference is to quantify the effect of X on Y. The gold standard for making causal inference is a randomized experiment in which the effect of X on Y can be isolated through the process of randomly assigning individuals to control and treatment groups. Successful randomization ensures equal distribution of underlying characteristics within the population such that the only difference between treatment and control groups is the intervention. Accordingly, any difference in the outcome may be attributed to an effect of the intervention or treatment. However, experimentation is often not financially or ethically possible, and scientists remain interested in testing hypotheses and inferring causation. Herein lies a challenge exemplified by the adage that correlation does not equal causation (Pearl & Mackenzie, 2018) – a challenge that was never more evident than when epidemiologists were confronted with understanding the causal effect of smoking on lung cancer (Proctor, 2012), and one that continues to plague analyses of observational data.

As shown in the fourth column of **Box 1**, a key distinguishing characteristic among these tasks is the need for causal knowledge on the relationship of interest in developing the analytical approach. Description and prediction do not necessarily require investigators to know the causal relationships among variables of interest, whereas such knowledge is relevant to associational analyses and foundational to causal inference. In the following section, we further delve into the task of causal inference in the context of EEB, and provide resources for those interested in using more complex statistical techniques (c.f., Young, Cain, Robins, O'Reilly, & Hernán, 2011).

## 4. THE ROADMAP: A BASIC GUIDE TO CAUSAL INFERENCE

Rarely, if ever, do construction workers build a house without reviewing and following an architect's blueprint, which is crafted to meet the needs of an occupant. Yet, scientists often overlook development of a data analysis plan (blueprint) prior to embarking on statistical analyses (construction work) to answer the research question (needs of the future occupant). The reasons for this oversight are multifaceted, ranging from lack of adequate training to research fatigue, the latter of which transpires from the fact that data analysis is preceded by an often, arduous period of data collection. Developing an analysis plan prior to implementing the analysis not only encourages articulation of the research question at hand, but also streamlines the analytical approach and prevents deviation from the original research goal (which in turn results in non-causal and inappropriate interpretations of model parameters).

Recently, interest in causal inference in EEB appears to be increasing. Causal diagrams are used to illustrate research questions and inform covariate adjustment (Laubach et al., 2020; Rosenbaum et al., 2020). Investigators use the word "effect" in reference to estimates derived from observational data (Laubach et al., 2019) in these fields. Thus, the goal of this section is to provide basic guidelines to making causal inference. **Figure 1** is a roadmap for steps to making causal inference, along with abridged suggestions for the other three tasks.

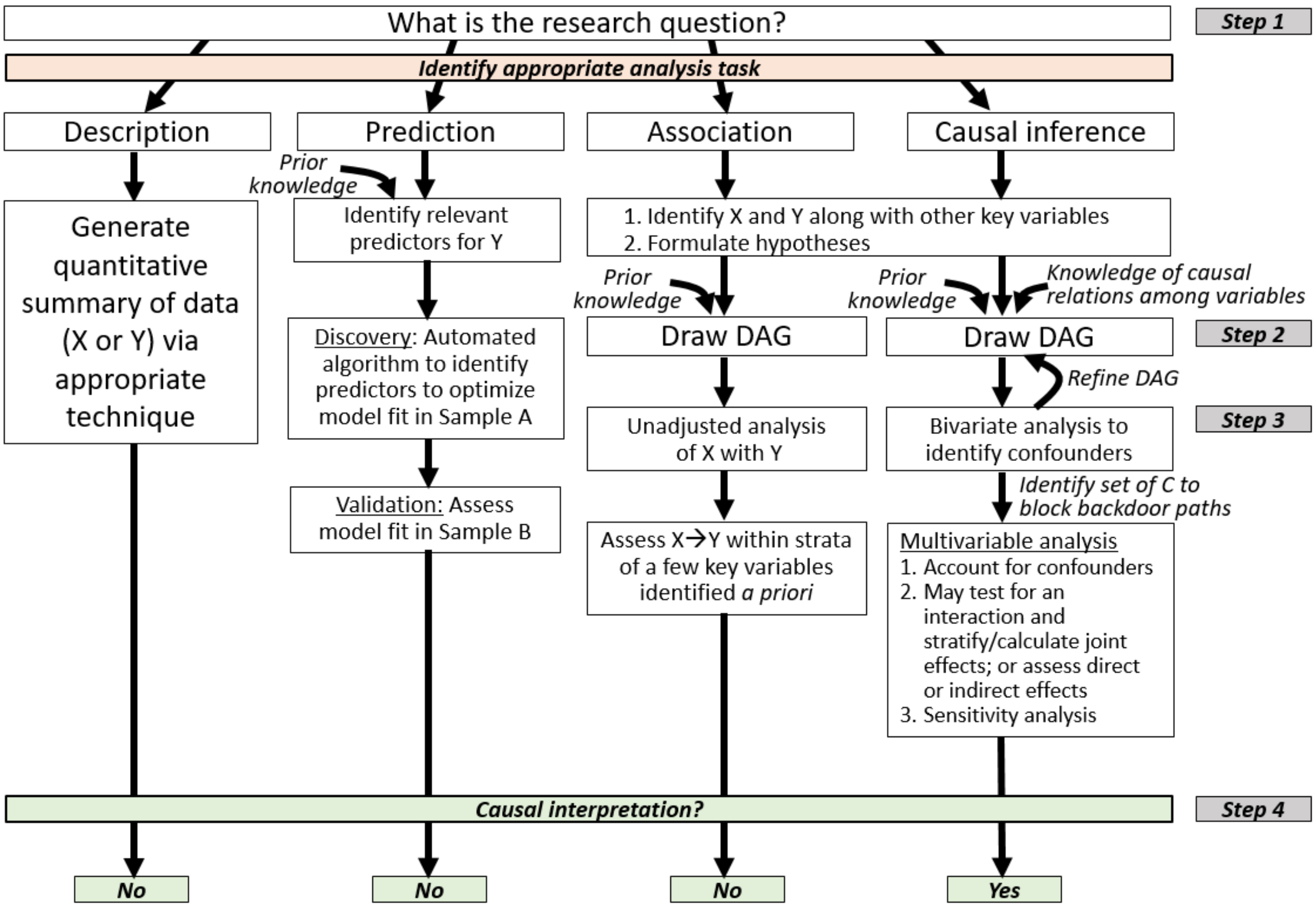

**FIGURE 1** Workflow to identify the appropriate data analysis task and modeling approach

### 4.1. Step 1. Formulate the research question.

Causal inference begins with a precise and specific research question (*nota bene*, this is true for all of the data analysis tasks, but especially for causal inference) which should be translatable into a hypothesis. Scientists from EEB are rigorously trained in development of testable hypotheses (Quinn & Dunham, 1983) and so we will not delve into it here. The requirements for causal inference include some notable extensions. A clear hypothesis for causal inference is one that can be translated into a contrast of counterfactual outcomes – that is to say, what Y would be had X been different in a fixed population. This requires clear identification of the X and Y of interest, specification of how X will be parametrized (i.e., discrete vs.

continuous) in order to obtain a consistent effect on Y (Hernán, 2016), and identification of the study population for whom data are available not only on X and Y, but also other key variables that warrant consideration in the analysis (confounders, precision mediators, effect modifiers, discussed below).

**4.2. Step 2. Draw the DAG**.

A causal directed acyclic graph (DAG) is a unidirectional flowchart that maps out the causal and temporal relationship between X and Y, while also considering third variables that may affect the X→Y association. As such, a DAG is a graphical representation of the hypothesis and a summary of the modeling approach (Greenland, Pearl, & Robins, 1999; Sauer & VanderWeele, 2013). Here, we provide an overview of DAGs, how they relate to the hypothesis, and describe their utility in development of an analytical approach. Throughout, we use an example hypothesis that social connectedness (measured via the number of affiliative interactions an individual has within their social network) affects immune function (T-cell count) in hyenas.

The first step to drawing a DAG is to identify the X (cause) and Y (effect) of interest. If our hypothesis is that social connectedness affects immune function, then the backbone of the DAG is an arrow emerging from X (social connectedness) pointing toward Y (immune function). This arrow aligns with the flow of time and the direction of causation (**Figure 2**), which translates into an unadjusted statistical model in which social connectedness is the explanatory variable and immune function is the dependent variable.

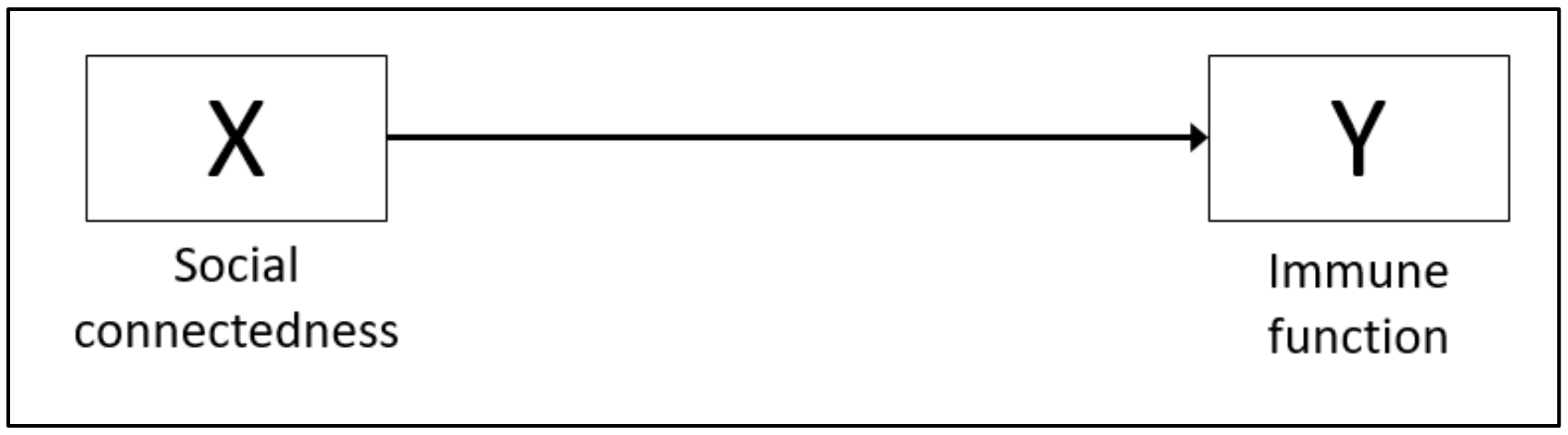

**FIGURE 2** DAG showing the relationship between X and Y.

The scenario shown in **Figure 2**, where X and Y are the only variables with which to contend is rare. In most natural systems, estimating the effect of X on Y requires consideration of other variables that can confound, modify, or mediate the relationship of interest.

### 4.2.1. Confounder

Upon defining the relationship of interest, the next step is to pinpoint potential sources of bias in the causal relationship. One source of bias is confounders, or shared common causes of X and Y. Confounders are identified via prior knowledge on the relationship of interest (i.e., known correlates or determinants of social connectedness and potential determinants of immune function) in conjunction with evidence of a statistical association of C with X and Y in the study sample, and should typically be accounted for in order to obtain a causal effect of X on Y.

In a DAG, a confounder is denoted by C with two arrows pointing at X and Y (**Figure 3**).

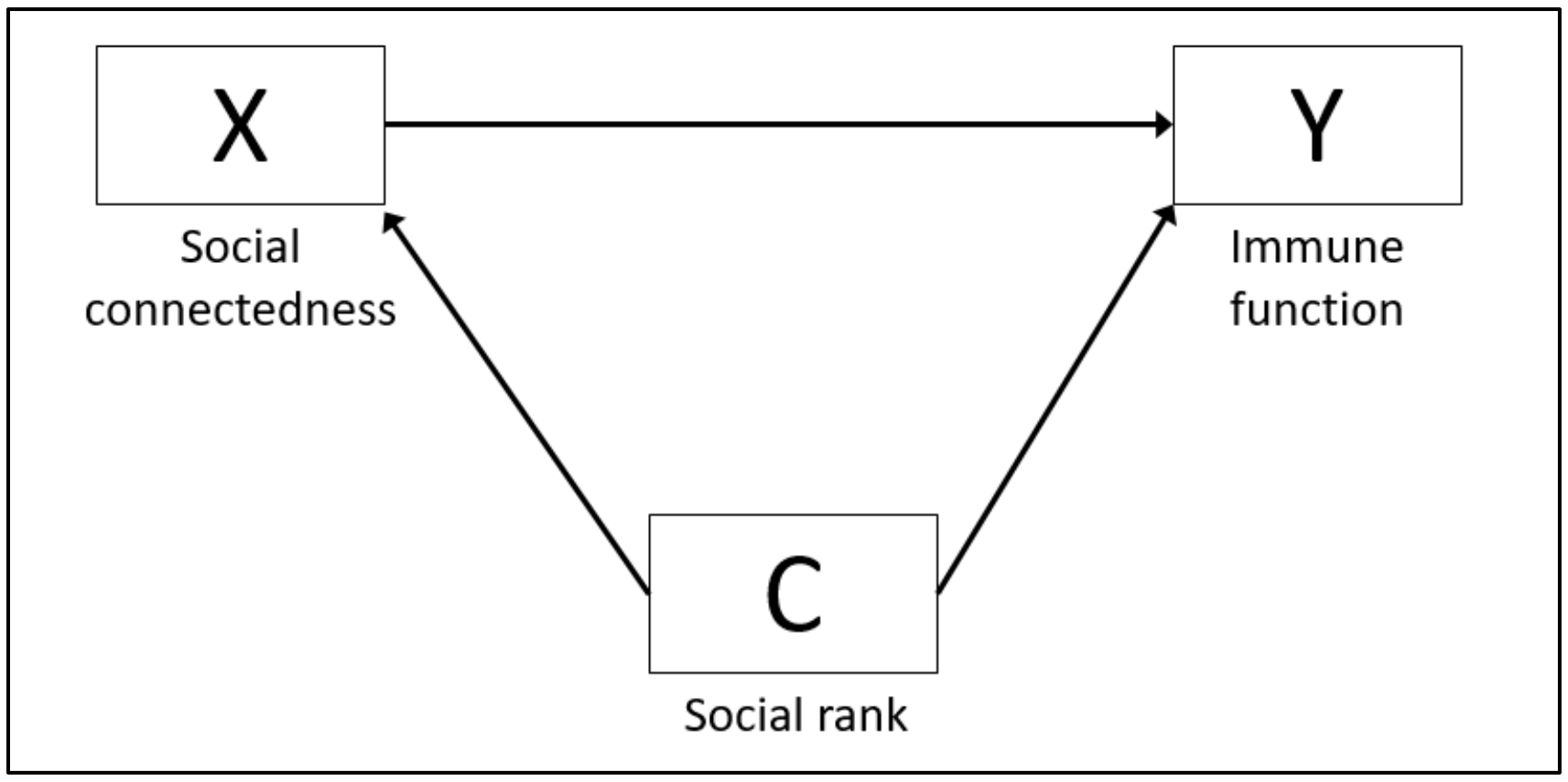

**FIGURE 3** DAG showing the relationship between X and Y, with confounder C.

When a confounder is present, there are two paths through which the effect of X may flow to Y: from X directly towards Y, and through a "backdoor path" that represents variation (and thus a potential non-causal association) in the relationship between X and Y transpiring from the shared common cause, C. In a causal analysis, the investigator is interested in the effect represented by the first arrow. Thus, when "open" or "unblocked," the backdoor path through C leads to confounding and should be "blocked" in order to obtain a causal estimate of X. Neglecting to do so may lead to biased estimates of association, or worse, spurious associations (Greenland & Pearl, 2017). Blocking a backdoor path (if not done at the phase of designing the study) can occur through statistical adjustment, stratifying, or weighting.

For instance, it is known that social rank affects both social connectedness (higher ranking individuals have more social connections) and immune function (higher ranking individuals have better access to food and resources, and thus better immune function), and therefore, is a possible confounder. The most common approach to control for confounding is to include the variable as a covariate in a regression model (adjustment). By including C as a covariate, the relationship between X and Y is theoretically "free" (independent) of its effect. While this approach is common, it is prone to fallacies that could introduce rather than reduce bias. Examples include use of the incorrect mathematical form of covariates (e.g., entering a covariate as a continuous variable when it is not linearly associated with the outcome), heterogeneity of effects of X on Y across levels of the covariate (i.e., existence of statistical interactions, discussed later), and collinearity with other covariates.

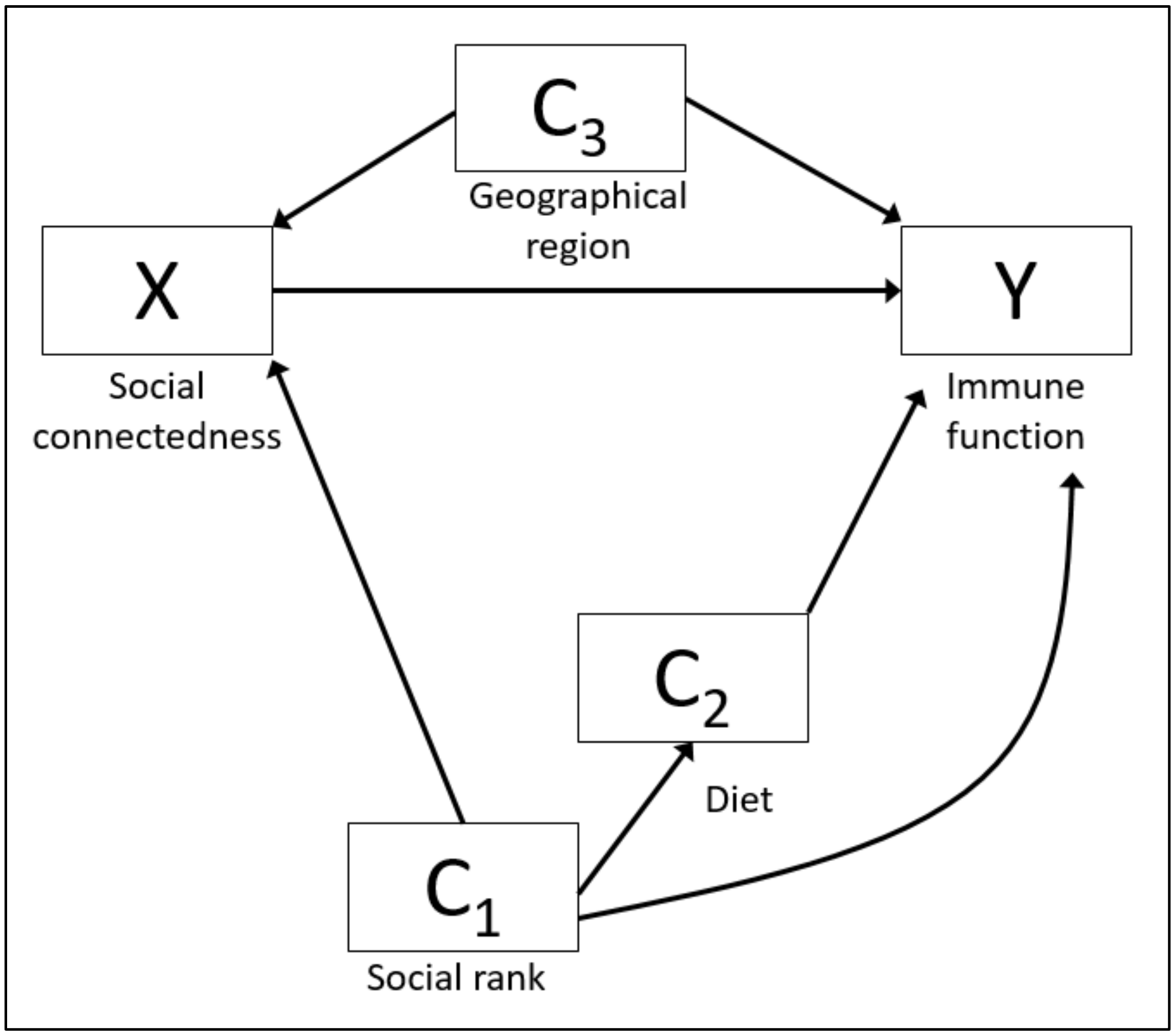

**FIGURE 4** DAG showing the relationship between X and Y and multiple confounders, $C_1$, $C_2$, and S.

It is worth nothing that for any given X→Y relationship, there are likely multiple confounders (**Figure 4**). Adjusting for all of them may not be feasible from a computational standpoint and may not be required from a causal inference standpoint. The analytic plan should be constructed so that all confounding (backdoor) paths are blocked. To do this, the investigator should map out the causal and temporal relationships among the confounders (tools like *daggity* [http://www.dagitty.net/] can help) to identify the most parsimonious set of variables to account for in order to block backdoor paths, as well as variables that should not be conditioned on in an analysis (see later section on Colliders). When multiple confounders lie on a shared backdoor path, there are decision-making points to progress through before arriving at the final model. In **Figure 4**, there are four paths from X to Y: the direct X → Y path, the backdoor path X → $C_1$ → $C_2$ → Y, the backdoor path X → $C_1$ → Y, and the backdoor path from X → $C_3$ → Y. Assuming that this DAG represents the true underlying relationships among all variables, only $C_1$ and $C_3$ need to be accounted for in the analysis in order to block backdoor paths. The choice of blocking the path through $C_3$ is simple as it lies on a backdoor path with no alternate routes. The choice between $C_1$ vs. $C_2$ is more complicated. Based on **Figure 4**, one should adjust for $C_1$ for two main reasons. First, blocking $C_2$ is insufficient because it leaves the X → $C_1$ → Y path open, allowing for a non-causal effect to flow from X to Y. Second, specific to this example, social rank is based on wins and losses from agonistic interactions in wild spotted hyenas and therefore is an objective metric that can be measured with minimal error. On the other hand, diet is difficult to measure since a researcher cannot be certain that s/he observed all instances of a hyena eating, how much it ate, or that the source of food was accurately identified. Thus, adjusting for a confounder derived from higher quality data is preferable.

Beyond statistical adjustment to account for confounders, other options include stratification (i.e., analysis of the X→Y relationship within subsamples of the population according to levels of C, though stratification aligns more with the concept of effect modification, discussed later) or inverse probability weighting (i.e., up-weighting or down-weighting the effect of a confounder based on its distribution in the study sample), which circumvents some assumptions required of statistical adjustment (Ellis & Brookhart, 2013; Gruber, Logan, Jarrín, Monge, & Hernán, 2015; Raad, Cornelius, Chan, Williamson, & Cro, 2020).

#### 4.2.2. Precision covariate

Often confused with confounders are precision covariates, or variables that are associated with X only or Y only.

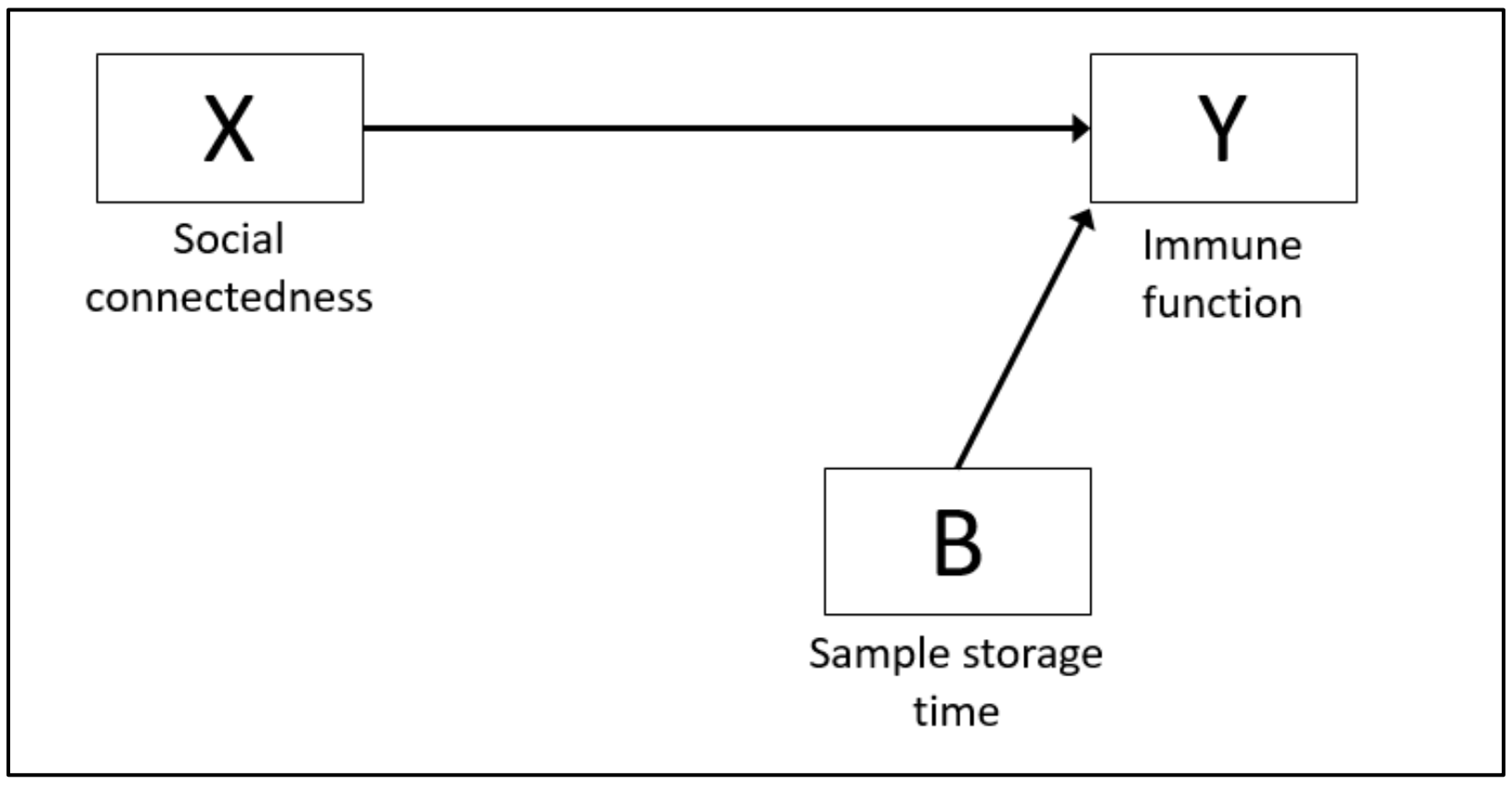

**FIGURE 5** DAG showing the relationship between X and Y and precision covariate B.

Precision covariates are denoted as B pointing towards X or Y (**Figure 5**), with no backdoor path linking X to Y. Such variables are typically used to account for technical variability and controlling for them in an analysis can improve model efficiency (Schisterman, Cole, & Platf, 2009). For instance, many analytes degrade over time even under the most pristine storage condition so biospecimen storage time is often included as a precision covariate in biomarker analyses.

#### 4.2.3. Mediator

A researcher may be interested in the extent to which X affects Y after controlling for a variable on the causal pathway between X and Y, also known as a mediator, M. As shown in **Figure 6**, M is a consequence of X and a determinant of Y.

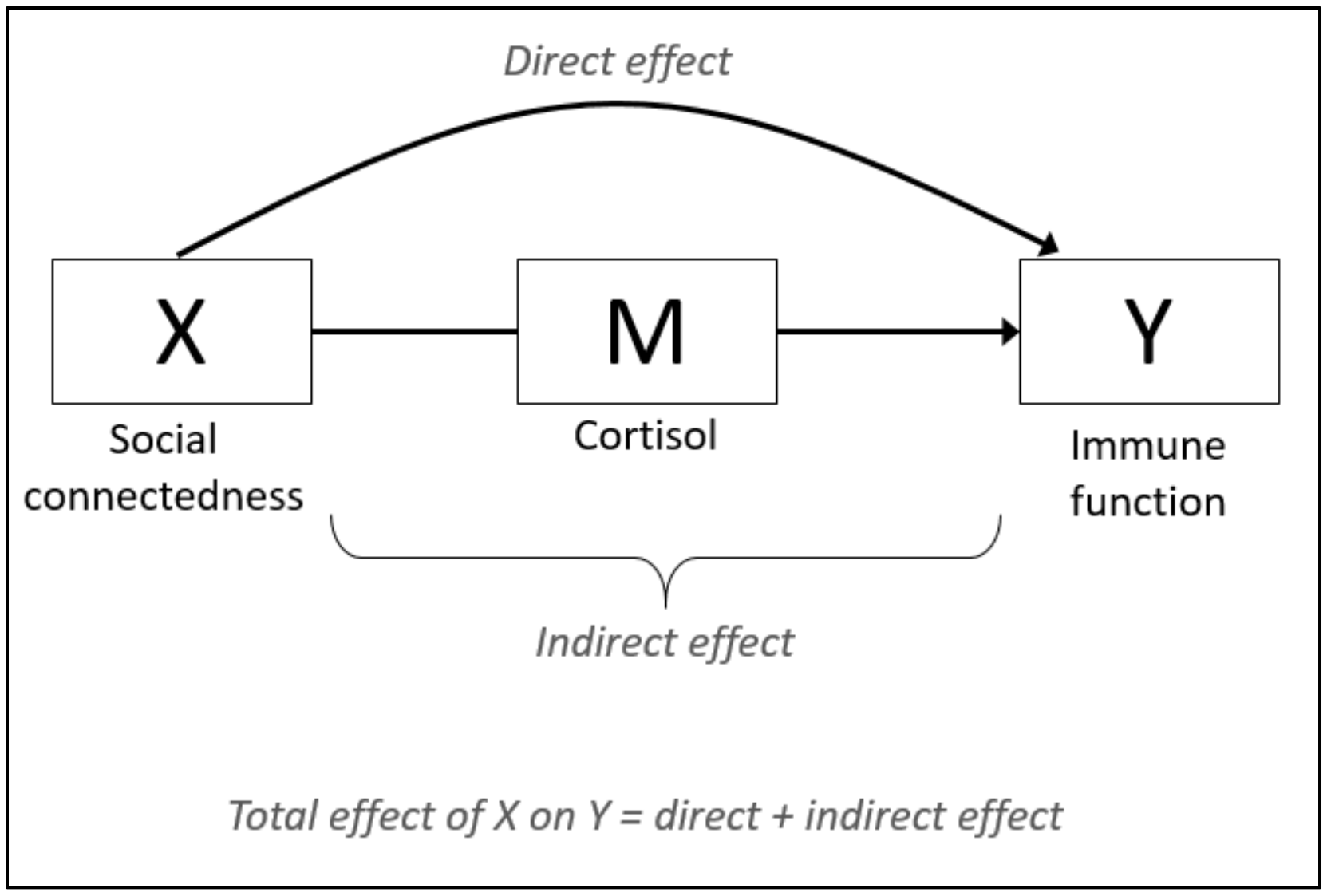

**FIGURE 6** DAG showing the relationship between X and Y, and mediator M.

Analyses that involve mediators are useful to assess possible mechanisms or pathways linking X to Y. Say we are interested in the biological mechanism underlying the relationship between social connectedness and immune function. We suspect that chronic activation of hypothalamic-pituitary-adrenal (HPA) axis is a potential mechanism given literature on chronic stress and immunosuppression, and that circulating cortisol is a marker of this pathway (Segerstrom & Miller, 2004). This hypothesis may be tested using a regression-based approach after evaluating some assumptions: (1) that the relationship between X and Y exists, (2) that M is associated with both X and Y, (3) that there is no interaction between X and M (Baron & Kenny, 1986), and (4) that there are no common causes of M and Y other than X (VanderWeele, 2015). If testable assumptions are met, the analysis involves comparing the estimate for X before (total effect) vs. after (direct effect) including M as a covariate in the model. If the estimate for X becomes attenuated (i.e., smaller/closer to the null) after adjusting for M, then M might represent a mediating pathway.

This approach is widely used but limited in utility since violation of testable and non-testable assumptions can introduce major bias into the estimate of interest. To start, this approach necessitates homogeneity of effects of X on Y across all levels of M. In other words, there should be no interaction between X and M – i.e., regardless of how high or low an animal's cortisol levels is, the effect of social connectedness and immune function should be exactly the same. Empirically, this may be tested via a test for statistical interaction, but it is difficult to conclude that an interaction term with P = 0.06 represents homogeneity of effects while an interaction P = 0.05 does not. Second, this approach assumes no unmeasured confounders between the M and Y. This heroic assumption cannot be tested empirically and if violated, can induce a specific type of bias (collider stratification bias) by conditioning on a variable caused by X and caused by a shared common cause of M and Y (Ananth & Schisterman, 2017). Doing so can result in a paradoxical flip in the direction of the estimate for X after conditioning on M. Third, this approach assumes that the measured variable M is a perfect proxy for the true mediator, which is rarely the case. In our example, cortisol is just one marker of HPA function and certainly does not capture the entire HPA cascade. Imperfect proxies for M lead to an underestimation of the mediating pathway (indirect effect), and an overestimation of the direct effect of X (Schisterman et al., 2009); such biases carry consequences for interpretation of results. There are now a number of non-parametric methods that relax or circumvent some of these assumptions discussed elsewhere (Coffman & Zhong, 2012; Zeng et al., 2020).

#### 4.2.4. Effect modification

If we have reason to believe that the relationship between social connectedness and immune function differs by a third variable that is an independent determinant of Y, then our hypothesis involves effect modification. In our example, the same degree of social connectedness for males vs. females may elicit differential effects on immune function (Olsen & Kovacs, 1996; Verthelyi, 2001). An effect modifier (Q) changes the nature of the relationship between X and Y (i.e., heterogeneity in the effect of X on Y across levels of Q (VanderWeele & Knol, 2014)) and is sometimes denoted with an arrow pointing towards the arrow between X and Y (**Figure 7**), though from a purely causal standpoint, Q is simply an independent determinant of Y.

Note that because DAGs are non-parametric, the potential for effect modification exists whenever two or more variables affect the same variable, and effect modification is bidirectional – that is, if Q modifies the effect of X on Y, then X also modifies the effect of Q on Y. Despite the bidirectionality of effect modification, a researcher should not lose sight of their question and should focus inference on the explanatory and outcome variables of interest.

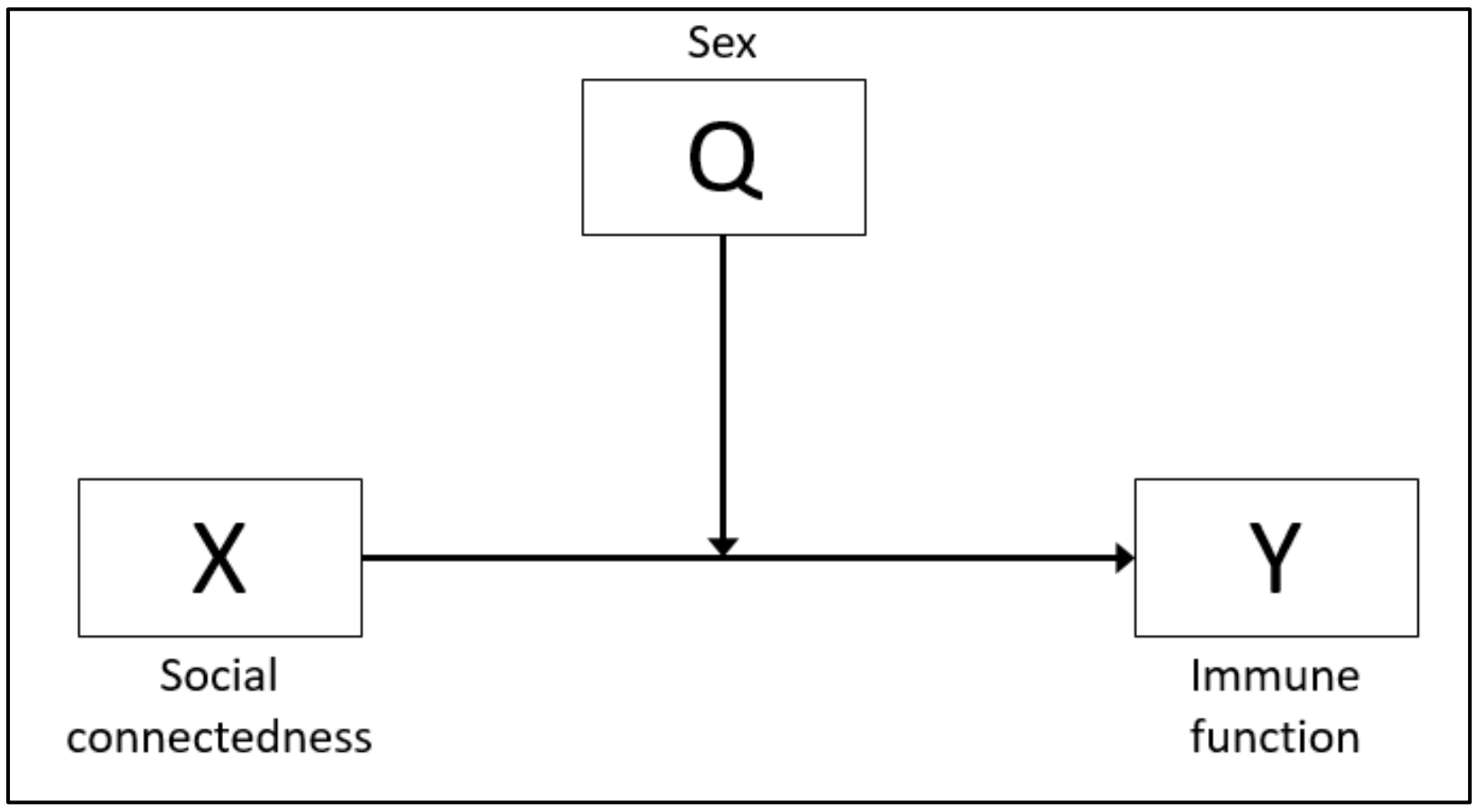

**FIGURE 7** DAG showing the relationship between X and Y and effect modifier Q.

The existence of effect modification may be assessed empirically via an interaction term between X and Q. If the P-value for the interaction term is significant at a predefined alpha, then effect modification is present. It is important to keep in mind that unlike confounding, effect modification is an intrinsic phenomenon – a form of biological truth – that should be observed rather than controlled for or “adjusted away” in the analysis. A parsimonious approach to effect modification is to test for a statistical interaction between X and Q, then stratify the analysis by levels of Q if appropriate.

Upon reading this, many statisticians ask, “Why stratify as opposed to interpret the estimate for the interaction term? After all, stratification decreases statistical power.” While one may certainly interpret the estimate for the interaction term, this value only provides information on the difference in magnitude of the X→Y relationship for one level of Q relative to another, but it does not tell you whether or how the direction of associations may differ for males vs. females (see Szklo & Nieto, 2019 for a more in-depth discussion of qualitative vs. quantitative interaction). To illustrate, the interaction term β-estimate for a linear regression model with a two-level effect modifier (Q = 1 for females, Q = 0 for males) would be the same if the estimates in stratified analyses were $\beta_{Q0} = -2$ and $\beta_{Q1} = 2$, as for $\beta_{Q0} = 2$ and $\beta = 6_{Q1}$. However, the fact that the estimates are in opposite directions for the former is meaningful and likely biologically relevant. Thus, noting the estimate for the interaction term itself, as well as exploring stratum-specific estimates with respect to the effect modifier are important to consider.

Of note, the concepts of statistical interaction, effect modification, and biological interaction are often conflated. Statistical interaction refers exclusively to the statistical significance of interaction term used to assess whether the effect of X on Y differs across levels of Q, and consequently, whether the investigator should conduct stratified analysis with respect to Q. Given a significant test for interaction, there may be effect modification or biological interaction. Though effect modification is often used as a catch-all term for heterogeneity of effects, some researchers posit that effect modification refers to heterogeneity of effects without synergism or antagonism between X and Q (VanderWeele & Knol, 2014). On the other hand, biological interaction refers specifically to synergism and antagonism (VanderWeele & Knol, 2014). A significant test for statistical interaction indicates the presence of either effect modification or biological interaction but does not tell you which. Moreover, whether a significant P-value for the X*Q interaction arises from effect modification or biological interaction cannot always be untangled empirically; the interested reader can learn more here (VanderWeele & Knol, 2014).

### 4.2.5. Collider

A collider (S) is associated both with X and Y. However, unlike confounders, colliders are a shared effect of X and Y (or the consequence of a shared effect of X and Y). This is depicted by arrows emerging from X and Y toward S. One should never condition on (i.e., adjust for, stratify by, or conduct subsample analyses with respect to) a collider; doing so will induce a spurious association between X and Y.

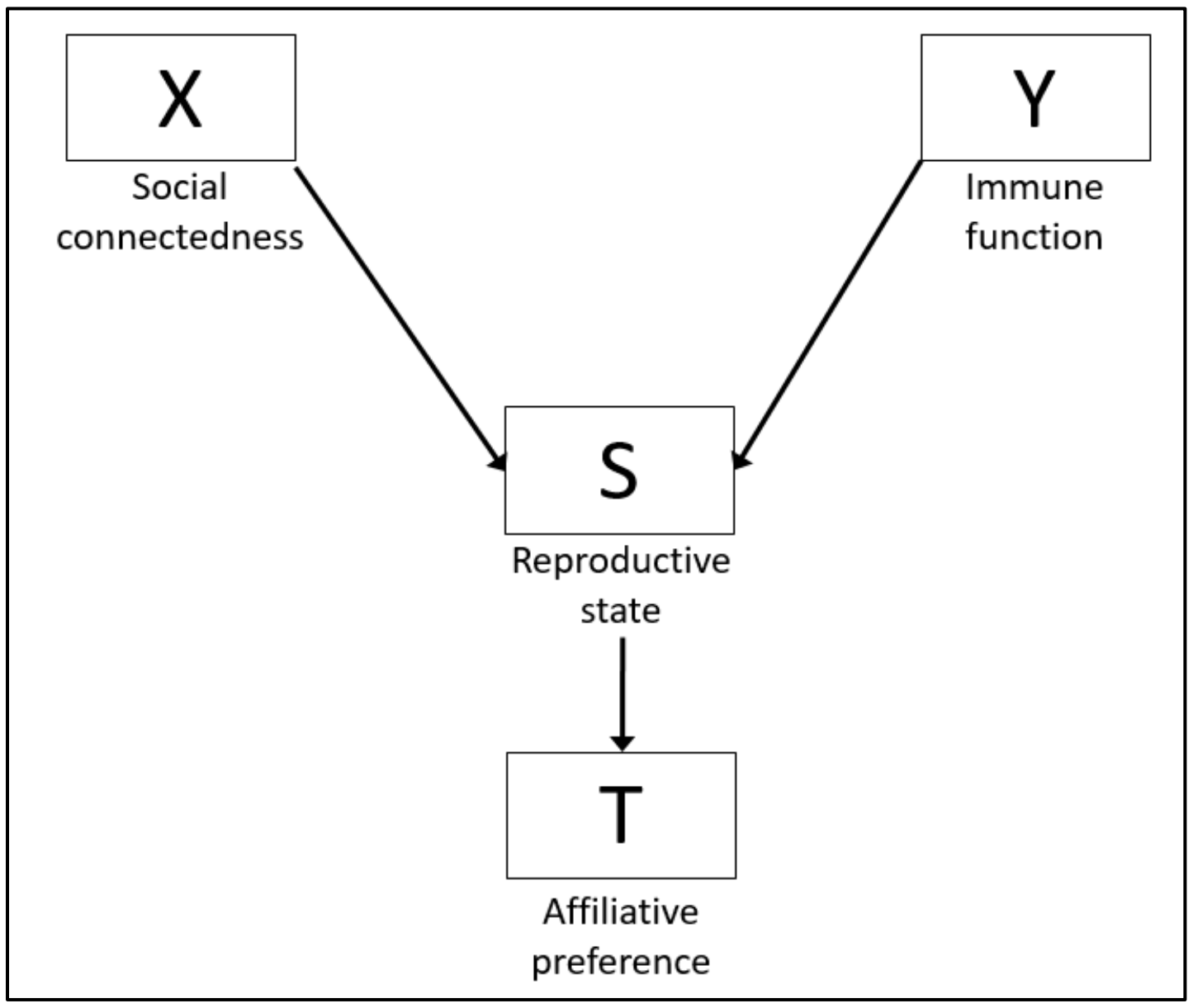

**FIGURE 8** DAG showing the relationship between X and Y, collider S, and a consequence of the collider T.

If, as shown in **Figure 8**, there is no association between X and Y (as denoted by the lack of arrow between the two), then conditioning on S (the collider) or T (a consequence of the collider) will force an association that otherwise does not exist. If an association does exist between X and Y, then conditioning on S can flip the estimate for X in the opposite direction. In our example, reproductive state – whether or not an animal is pregnant – may be a collider since aspects of the social environment, like social connectedness may influence reproductive state (higher social connectedness may be related to greater likelihood of being pregnant) and immune function may also affect reproductive state (dysregulated immune function is associated failed pregnancy). In this scenario, conditioning on reproductive state or affiliative preference due to reproductive state (a pregnant female may be less interested in affiliating with males than when she is not pregnant) forces us to observe the correlation between social connectedness and immune function only among individuals who are not pregnant, and only among those who are pregnant. Among pregnant individuals, we will observe a strong positive association between social connectedness and immune function, purely due to the fact that we conditioned on a collider. This phenomenon is troubling and has puzzled perinatal epidemiologists who erroneously conditioned on timing of parturition as a confounder in analyses where it was actually a collider due to unmeasured confounding (Ananth & Schisterman, 2017). Having encountered both confounders and colliders, we emphasize the utility of expert knowledge and DAGs to conceptualize both measured and unmeasured variables that may affect the relationship of interest, as opposed to relying solely on statistical associations (VanderWeele & Robins, 2007; Howards, Schisterman, Poole, Kaufman, & Weinberg, 2012).

### 4.3. Step 3. Implement the analysis.

#### 4.3.1. Univariate analysis & descriptive statistics

Prior to formal analyses, the researcher should familiarize themselves with the data (c.f. Prabhakaran, 2017; Zuur, Ieno, & Smith, 2007) via frequency distributions for continuous variables and contingency tables for categorical variables. These summary statistics provide a basic understanding of central tendency and spread, allowing the reader to assess the biological relevance of effect sizes in downstream analyses. They are also instrumental to identifying errors, missing values, outliers, and provide information on the distribution of the data. Centering and scaling variables should also take place at this time to enhance interpretation of estimates.

#### 4.3.2. Bivariate associations

Next, the investigator should conduct bivariate analyses to assess raw associations between potential confounders and precision covariates with the X and/or Y variables. In addition to covariate selection, bivariate associations allow for assessment of the shape of association between pairs of variables (linear, quadratic, polynomial). If an investigator identifies non-linear associations, then s/he should consider operationalizing variables to reflect this. The simplest approach is to enter the variable in a non-parametric format, such as in quantiles, to allow for non-linear associations. At this point in time, the investigator may modify their DAG based on statistical associations observed in the dataset. For instance, if a variable was initially included as a confounder in the DAG but is not associated with X or Y in the study sample, then the investigator may decide to remove it from the DAG.

### 4.3.3. Main analysis

The main analysis should reflect the DAG and may contain multiple elements from **Figures 2-8**. While the number of research questions is infinite, there are typically two broad endpoints for a given research question: quantifying the total effect of X on Y, and quantifying direct vs. indirect effects (mediation analysis) of X on Y.

*Total effects*

If interested in the total effect of X on Y, the investigator should not adjust for mediators since mediators are part of the effect of X. Accounting for a mediator when the goal of an analysis is to estimate total effects results in null-biased estimates (Schisterman et al., 2009). This does not imply that the investigator should avoid adjusting for confounders. As a shared common cause of X and Y, accounting for a confounder does not block any of the total effect of X but rather, controls for extraneous variability in the relationship between X and Y. Precision covariates may also be considered at this point in time.

When selecting covariates for the multiple variable models, controlling bias is the goal, followed by parsimony. After identifying possible confounders via a DAG and bivariate analysis, it is prudent to identify the smallest set of confounders to account for in order to block all backdoor paths from X to Y. When the need to adjust for a variable is in question, we suggest examining the association of interest with vs. without adjustment for a confounder or precision covariate; if including a variable does not appreciably change the results, then it is preferable to proceed without adjustment (Schisterman et al., 2009).

*Direct and indirect effects*

If a research question involves assessing the involvement of a biological mechanism or pathways, then the analysis involves estimation of direct and indirect effects. While we provide some guidance here, this type of analysis requires several assumptions, only some of which are testable, and is more complex than the scope of this commentary. Researchers commonly employ the previously-mentioned Barron and Kenny's approach (Baron & Kenny, 1986). As discussed earlier, there are limitations to this approach, and caution should be taken when interpreting direct and indirect effects given the assumptions required and because the variables for which we have data are likely upstream or downstream proxies of the true mediator or possibly even markers of a parallel physiological pathway. As an over-simplified example, a naïve investigator might test blood glucose levels as a mediator to the relationship between social connectedness and immune function and conclude that glycemic regulation is the key mechanism linking social connectedness to immune function. Yet, elevated circulating glucose occurs secondarily to elevations in cortisol, the latter of which is the true culprit. In such instances, not only will the direct and indirect effects be biased, but also, declarative conclusions about glycemic regulation as the mechanism underlying the relationship of interest will be partially false. In addition to being aware of ways in which mediation analyses go awry, the investigator should not claim that the mechanism of interest is responsible for the effect of X on Y. Rather, the s/he should acknowledge that the biomarker is purported to represent the mechanism of interest, and that subsequent *in vivo* or *in vitro* studies are required to interrogate mechanisms identified from observational studies. Investigators should be particularly careful when interrogating mediators that cannot be directly manipulated, as these variables are particularly unlikely to meet the required assumptions (Murray, Robins, Seage, Freedberg, & Hernán, 2020).

*What about effect modification?*

In analyses of total effects, assessment of statistical interactions and stratified analyses can be carried out after finalizing the multiple variable model, adjusting for the same set of covariates within strata of the effect modifier – so long as the effect modifier is not a collider.

In a mediation analysis that also involves effect modification, the above approach is generally reasonable if the effect modifier temporally precedes the exposure. One may assess the need to stratify based on the total effect of X on Y, then conduct regression-based mediation analysis within strata of the effect modifier. However, such an approach is not appropriate if the effect modifier occurs after the exposure and thus, is also a mediator (recall the assumption that there should be no interaction between X and M in order to implement Barron and Kenny's method). In this scenario, there are alternate approaches to mediation, such as the counterfactual based approach (Valeri & VanderWeele, 2013) which focuses on the natural direct effect of X by estimating an effect of X at a specific value or level of Q that is allowed to vary between subjects, the controlled direct effect of X by estimating an effect of X at a single value of Q for all subjects (Vanderweele, 2011), or principal stratum conditioning for discrete mediators (VanderWeele, 2011), which estimates the natural direct effect of X on Y only among individuals who experience the mediator (the principal stratum). Each of these approaches focus on estimating the effect of X on Y at a given value of Q, as opposed to taking the average across all levels of Q when heterogeneity of effects is suspected. While such approaches are theoretically sound, they can be problematic in practice as interpretation of such effects often do not reflect real life scenarios – i.e., assessing a direct effect of X at a single value or stratum of Q is an oversimplification of biology.

#### 4.3.4. Step 4. Interpret results

When interpreting the results, it is important to focus on the specific relationship between X and Y, and avoid temptations of interpreting estimates for other covariates in the model. When identifying noteworthy findings to discuss, avoid worshipping the P-value – instead, focus on the direction, magnitude, and precision of the estimates across sensitivity analyses since a truly robust association is likely to persist.

Keep in mind that for any given hypothesis, there are alternate hypotheses and thus, alternate explanations for findings. Acknowledge and search for these possibilities, whether they transpire from complementary biological phenomena, unmeasured confounders, or inappropriate statistical adjustment (e.g., adjusting baseline values of a variable when the outcome is change in that variable (Glymour, Weuve, Berkman, Kawachi, & Robins, 2005; Pearl, 2014), adjusting for a collider (Hernán, Clayton, & Keiding, 2011), or adjusting for a mediator that has a shared common cause with the outcome (Tu, West, Ellison, & Gilthorpe, 2005; Ananth & Schisterman, 2017). Ideally, an investigator would have assessed for and ruled out some of these possibilities via the formulation of the DAG and sensitivity analyses, though some find it helpful to revisit the DAG after the analysis to sketch out unmeasured variables that may be responsible for surprising or alternate findings. Presenting the DAG and justifying the choice of variables and arrows therein is a valuable way of conveying the necessary assumptions that must hold for the results to be interpreted causally (Tennant et al., 2019).

## 5. FOOD FOR (CAUSAL) THOUGHT

Although the four data analysis tasks discussed herein are self-contained analytical approaches, they do not act in isolation. Associational analyses can inform future studies of causal inference or prediction; knowledge of key associations can point toward causation between two specific variables, and/or lay the foundation for identification of variables that may be relevant as an explanatory variable in a predictive model. Likewise, assessment of variance and central tendencies (description) are almost always implemented at the start of all analyses in order to assess assumptions for multivariable normality and to identify errors in the data. The distinction among the tasks lies in the need for causal knowledge when building the models and the interpretation of results. No causal or temporal knowledge is necessary for descriptive, predictive, or associational analyses, and thus, interpretation of results from such analyses should not use causal language (i.e., no use of the word "effect"). On the other hand, causal inference seeks to quantify the specific relationship between X and Y by controlling for extraneous bias by confounders and by carefully considering other third variables that may influence the relationship of interest. When implemented carefully and thoughtfully, such analyses do seek to quantify an effect of X on Y. Also, worth mentioning is that even when investigators seek to make causal inference, they may not have the necessary/appropriate data or training to do so. In such a scenario, it is appropriate to begin with associational analyses in order to understand direction and magnitude of a relationship of interest via unadjusted associations (which also avoids the need to accept assumptions of parametric regression models) as a first step towards future studies of causal inference. In our opinion, studies that use description or association tasks warrant equal consideration for publication as compared to prediction and causal inference, given that they do not answer inferior questions but rather they provide insight on raw biological variation invoking far fewer assumptions.

## 6. CONCLUSIONS

Three years after pointing out the flaws of models, Box updated his outlook stating, "All models are wrong, but some are useful," (Box, 1979). In espousing Box's 1979 view, we hope that this commentary and the resources it points to will help others develop useful models to answer important biological questions, particularly when making causal inference with observational data is the goal.

Like the meticulous epidemiologist, cautious behavioral ecologists and evolutionary biologists are hesitant to directly relate results from an observational analysis to the research question that they set out to answer, adding confusion and semantics to biological interpretation – perhaps an artefact of stigma around use of causal language to describe observational findings (Pearl & Mackenzie, 2018). Through this commentary, we hope to have harnessed synergy between two historically distinct worlds of research in order to streamline the process of data analysis in a consistent and methodical manner.

Modern causal inference, and much of the basis for this paper, transpires from the work of torchbearers including but not limited to Judea Pearl, Sander Greenland, James Robins, Tyler VanderWeele, Miguel Hernán, Enrique F. Schisterman, Eric Tchetgen Tchetgen, Maria Glymour, Andrea Rotnitzky, and Jessica G. Young. While these names may be unfamiliar to EEB scientists, Sewall Green Wright (1889-1988), an American evolutionary biologist known for his work involving path analysis, is a household name and a direct ancestor to modern causal inference (Wright, 1921, 1960). Thus, while causal inference went out of fashion for much of the 20$^{th}$ century among EEB researchers, it is rooted to luminaries in the field from the early- to mid-19$^{th}$ century. In recent years, we have noticed an increasing number of EEB scientists who have developed an interest in causal inference and formed collaborations with epidemiologists and/or biostatisticians – a rich collaboration in which we hope to participate and promote.

**Funding:**

ZML is supported by the Morris Animal Foundation grant D19ZO-411. WP is supported by the Center for Clinical and Translational Sciences Institute KL2-TR002534. RJS is supported by the National Science Foundation, IOS 1856266. KLH is supported by National Science Foundation grant DEB-1911619.

**Author contributions:**

ZML and WP conceived the ideas and lead the writing of the manuscript; EJM, KLH, RJS provided content expertise in causal inference (EJM), as well as behavioral ecology and evolutionary biology (KLH and RJS). All authors contributed critically to the drafts and gave final approval for publication.

**Competing interests:**

Authors declare no competing interests.

**Data and materials availability:**

No empirical data were collected for this project.